\documentclass[amstex,amsfonts,twoside,
preprint,
aip,
amsmath,a4paper]{revtex4-1}
\usepackage{bm}
\usepackage{graphicx}
\usepackage{xcolor}
\usepackage[T1]{fontenc}
\usepackage{hyperref}

\begin{document}

\newcommand*\jr[1]{{\color{red} [#1]}}

\setcounter{page}{1}

\title{
One-Shot Simulation of Static Disorder in Quantum Dynamics
with Equilibrium Initial State via Matrix Product State
Sampling
}

\author{Zhao Zhang}

\author{Jiajun Ren}
\email{jjren@bnu.edu.cn}

\author{Wei-Hai Fang \vspace*{0.5em}}
\affiliation{\sffamily Key Laboratory of Theoretical and
Computational Photochemistry, Ministry of Education, College of
Chemistry, Beijing Normal University, Beijing 100875, People's
Republic of China}

\begin{abstract}

    Static disorder plays a crucial role in the electronic
    dynamics and spectroscopy of complex molecular systems.
    Traditionally, obtaining observables averaged over static
    disorder requires thousands of realizations via direct
    sampling of the disorder distribution, leading to high
    computational costs. In this work, we extend the auxiliary
    degree-of-freedom based matrix product state (MPS) method to
    handle system-bath correlated thermal equilibrium initial
    states. We validate the effectiveness of the extended method
    by computing the dipole-dipole time correlation function of
    the Holstein model relevant to the emission spectrum of
    molecular aggregates. Our results show that the method
    accurately captures static disorder effects using a
    one-shot quantum dynamical simulation, with only a
    moderate increase in MPS bond dimension, thereby
    significantly reducing computational cost. Moreover, it
    enables the generation of a much larger number of samples
    than the conventional direct sampling method at negligible
    additional cost, thus reducing statistical errors. This
    method provides a broadly useful tool for calculating
    equilibrium time correlation functions in system-bath
    coupled models with static disorder.

 
\end{abstract}
\maketitle

\section{INTRODUCTION}

Static disorder, arising from defects, structural fluctuations,
or packing imperfections, can have a profound influence on
electronic dynamics and spectroscopy in molecular aggregates and
materials, such as charge and exciton migration, transport, and
relaxation processes.~\cite{moix2013coherent,zhong2014coherent,
hestand2018expanded,peng2021influences,brey2024coherent,
zheng2020study,zheng2019charge,chaillet2020static,
li2025theoretical,zhang2025effect}
In theoretical modeling, static disorder is typically
represented by distributions in local site energies or
electronic couplings, often assumed to follow a Gaussian
distribution.~\cite{bassler1993charge,jang2001characterization,
hestand2018expanded,makri2024path} To
calculate observables averaged over static disorder, one
typically sample a large number of disorder realizations and
perform independent simulations for each. When each realization
involves computationally intensive simulations, such as full
quantum dynamics, the overall cost becomes prohibitively high,
even though these simulations are perfectly parallelizable.

To address this challenge, in 2021, Gelin \textit{et al.}
proposed an auxiliary degree-of-freedom (DoF) approach that
incorporates static disorder into a single quantum dynamical
simulation,~\cite{gelin2021efficient} thereby eliminating the
need for averaging over thousands of realizations. When combined
with efficient time-dependent matrix product state (MPS)
algorithms,~\cite{schollwock2011density} this method has been successfully applied to
simulate static disorder effects in charge transfer dynamics
within the Holstein-Peierls model,~\cite{gelin2021efficient} and energy
transport in the eight-site Fenna-Matthews-Olson
model.~\cite{sheng2024td} However, the original
auxiliary-DoF-based approach cannot handle system-bath
correlated initial state at thermal equilibrium, which depends
on the static disorder itself. Such a correlated equilibrium state
are crucial for calculating equilibrium time correlation
functions and dynamical response properties in system-bath
coupled systems.~\cite{nitzan2024chemical}

To overcome this limitation, in this work, we propose a new
MPS-sampling algorithm that extends the auxiliary-DoF method to
handle static-disorder-dependent thermal equilibrium initial
states. We demonstrate the validity of our approach by
calculating the dipole-dipole time correlation function of the
Holstein model under varying disorder strengths and system
sizes. Our method achieves comparable accuracy to the standard
direct sampling approach, but with only a one-shot simulation,
leading to significant savings in computational cost.

The remainder of this paper is organized as follows.
Section~\ref{sec:Method} presents the detailed algorithm.
Section~\ref{sec:Results} provides numerical examples
comparing our MPS-sampling approach with the traditional direct
sampling approach. Finally, conclusions are summarized in
Section~\ref{sec:Conclusion}.

\section{Method} \label{sec:Method}

\subsection{static disorder}

We focus on the calculation of quantum two-time correlation
functions (TCFs) $\langle \hat A(t) B \rangle$. This quantity
reduces to the time-dependent expectation value of a single
observable $\langle \hat A(t)\rangle$ when $\hat B = \hat I$,
and it can be naturally generalized to higher-order correlation
functions.

In the presence of static disorder, the TCF becomes
\begin{gather}
    \langle \hat A(t) \hat B \rangle = \int  \langle \hat
    A(\mathbf s, t) \hat B(\mathbf s)
    \rangle \rho(\mathbf s) d\mathbf s. \label{eq:tcf}
\end{gather}
For a pure initial state,
\begin{gather}
    \langle \hat A(\mathbf s,t) \hat B(\mathbf s) \rangle = \langle \psi
    (\mathbf s,0)|
    e^{i\hat H(\mathbf s) t} \hat A(\mathbf s) e^{-i\hat
    H(\mathbf s) t} \hat B(\mathbf s)|\psi (\mathbf s,0)\rangle
    \label{eq:ZT}
\end{gather}
For a thermal equilibrium initial state,
\begin{gather}
    \langle \hat A(\mathbf s,t) \hat B(\mathbf s) \rangle  =
    \textrm{Tr} [
    \frac{e^{-\beta \hat H(\mathbf s)}}{Z(\mathbf s)} e^{i\hat
    H(\mathbf s) t} \hat A(\mathbf s)
    e^{-i\hat H(\mathbf s)t} \hat B(\mathbf s)]. \label{eq:FT}
\end{gather}
$\mathbf s = \{s_1, s_2,\cdots\}$ represents the parameters
characterizing static disorder (e.g., fluctuation in site
energies), and $\rho(\mathbf s)$ is their probability
distribution, satisfying $\rho(\mathbf{s}) > 0$ and $\int
\rho(\mathbf{s}) d\mathbf{s} = 1$. The Hamiltonian $\hat{H}$,
the observables $\hat{A}$ and $\hat{B}$, and the initial state
all depend on $\mathbf{s}$ in general.

The standard approach to calculate Eq.~\eqref{eq:tcf} is through
Monte Carlo sampling of $\mathbf{s}$ according to
$\rho(\mathbf{s})$. 
\begin{gather}
    \langle \hat A(t) \hat B \rangle \approx \frac{1}
    {N_\textrm{samp}} \sum_{k=1}^{N_\textrm{samp}}
    \langle \hat A(\mathbf s_k, t) \hat B(\mathbf s_k)\rangle
\end{gather}
We refer to this as the direct sampling
method. It requires a large number of independent quantum
dynamics simulations, leading to high computational cost. The
statistical error in this method scales as
$1/\sqrt{N_{\textrm{samp}}}$, where $N_{\textrm{samp}}$ is
the number of samples.

\subsection{auxiliary-DoF-based method with
static-disorder-independent initial states}

To reduce the high computational cost of direct sampling, Gelin
\textit{et al.} proposed a novel algorithm in which auxiliary
$\mathbf{s}$ are introduced into the quantum
system.~\cite{gelin2021efficient} The wavefunction of the
auxiliary DoF $\chi(\mathbf{s})$ is chosen such that
$\chi^*(\mathbf{s}) \chi(\mathbf{s}) = \rho(\mathbf{s}) $,
allowing the effect of disorder to be encoded into a single
simulation.

In the original formulation, $\rho(\mathbf{s})$ was assumed to
be Gaussian, 
\begin{gather}
    \rho(\mathbf s) = \prod_i \rho(s_i) \\
    \rho(s_i) = \frac{1}{\sqrt{2\pi}\sigma}e^{-
    s_i^2 /2\sigma^2}
\end{gather}
Thus, $\chi(\mathbf{s})$ is naturally the ground
state wavefunction of harmonic oscillators, which is 
\begin{gather}
    \chi(\mathbf s) = \prod_i \chi_0(s_i) \\
    \chi_0(s_i) = (\frac{\omega}{\pi})^{1/4}e^{-\frac{1}{2}
    (s_i^2/(1/\omega))}.
\end{gather}
To satisfy $|\chi_0(s_i)|^2 = \rho(s_i)$, the frequency of the
oscillator is chosen as $\omega = 1/(2\sigma^2)$.

In this work, we generalize this formulation to arbitrary static
disorder distributions by working in the ``coordinate space''
rather than the bosonic Fock space used in the original method.
We define the auxiliary wavefunction as
\begin{gather}
    | \chi \rangle = \int d\mathbf s \sqrt{\rho(\mathbf s)} |
    \mathbf s \rangle.
\end{gather}
Accordingly, the static disorder related parameters in the
Hamiltonian becomes operators of the auxiliary DoFs, which
behaves like coordinate operators and fullfills
\begin{gather}
    \hat {s}_i |s_i \rangle = s_i | s_i\rangle.
\end{gather}
For example, to describe the static energetic disorder of local
electronic state,
\begin{gather}
    \hat H(\mathbf s) = \sum_i (\varepsilon_0 + s_i)  |\phi_i\rangle \langle
    \phi_i|, s_i \sim \rho_i(s_i)  \\
    \to \hat H(\hat {\mathbf s}) = \sum_i (\varepsilon_0 + \hat s_i)
    |\phi_i\rangle \langle \phi_i |.
\end{gather}

If the initial state is independent of static disorder, i.e.,
$\psi(\mathbf{s}, 0) \equiv \psi(0)$, Eq.~\eqref{eq:tcf}
simplifies to
\begin{gather}
    \langle \hat A(t)B \rangle = \langle \chi |
    \langle \psi(0) |\hat
    A(\hat{\mathbf s}, t) \hat B(\hat{\mathbf s})| \psi (0)\rangle |
    \chi \rangle \\
    |\chi \rangle = \prod_i |\chi_i\rangle 
\end{gather}
With this auxiliary-DoF-based method, only one single quantum
dynamics simulation from initial state $|\Psi(0)\rangle =
|\psi(0)\rangle|\chi \rangle$ is needed to obtain the
static-disorder-averaged quantity, with the trade-off being the
inclusion of additional auxiliary DoFs, which enlarge the
Hilbert space.

In practice, the continuous variables $\hat{\mathbf{s}}$ must be
discretized. This can be done using either a harmonic oscillator
basis (as in Ref.~\cite{gelin2021efficient,sheng2024td}) or a
discrete variable representation (DVR),
~\cite{light1985generalized,light2000discrete} as employed in
this work. With a sufficiently large basis size, the results
converge to the continuum limit.
The auxiliary-DoF-based method can be effectively combined with
time-dependent algorithms based on tensor network states, such
as time-dependent density matrix renormalization group (TD-DMRG)
~\cite{vidal2004efficient,daley2004time,feiguin2005time,ronca2017time,ma2018time,ren2022time,ma2022density,borrelli2021finite} or multilayer multi-configurational time-dependent
Hartree (ML-MCTDH)~\cite{wang2003multilayer,manthe2008multilayer,vendrell2011multilayer,wang2015multilayer,beck2000multiconfiguration}. The additional computational cost
due to auxiliary DoFs is expected to be negligible. Here, we
discuss qualitatively when this expectation is reasonable. In
the auxiliary-DoF-based method, the overall time-evolved
wavefunction $\Psi(t)$ is actually a linear combination of all
the wavefunctions each with specific static disorder $\mathbf s$
weighted by $\sqrt{\rho(\mathbf s)}$ 
\begin{gather}
    \Psi(t) = \int d \mathbf s \sqrt{\rho(\mathbf s)}
     |\psi(\mathbf s,t)\rangle |\mathbf s\rangle \label{eq:linearcombine}\\
    \psi(\mathbf s,t) =  e^{-i\hat H(\mathbf s) t} \psi(0)
\end{gather}
If the individual wavefunctions $|\psi(\mathbf{s}, t)\rangle$
are similar across $\mathbf{s}$, then the combined wavefunction
can be efficiently compressed by tensor network algorithms. This
condition holds when the static disorder is moderate, and thus
the disorder does not significantly alter the wavefunction.

Despite its appeal, this method is limited to
static-disorder-independent initial states. For thermal
correlation functions starting from a system-bath correlated
equilibrium state (Eq.~\eqref{eq:FT}), the initial state depends
on $\mathbf{s}$ itself. To overcome this, we develop a new
auxiliary-DoF-based MPS-sampling method.

\subsection{auxiliary-DoF-based MPS-sampling method with
static-disorder-dependent thermal equilibrium initial states}

We now describe how to extend the auxiliary-DoF approach to
thermal equilibrium initial states that depend on static
disorder, using the purification method in combination with
sampling MPS.

In the purification method, an auxiliary space $Q$ is introduced
(distinct from the auxiliary DoFs representing disorder). This
allows the mixed thermal state in the physical space $P$ to be
represented as a pure state $|\psi\rangle$ in the enlarged space
$P \otimes Q$, satisfying $\rho_P = \textrm{Tr}_Q |\psi\rangle
\langle \psi|$.~\cite{feiguin2005finite,verstraete2004matrix}

At infinite high temperature ($\beta=0$), the purified thermal
state $|\psi_{\beta=0} \rangle$ is trivially known, 
\begin{gather}
    |\psi_{\beta=0}\rangle = \sum_{\{\sigma,\tilde{\sigma}\}} \frac{1}{\sqrt{d^N}} 
    |\sigma_1 \tilde{\sigma}_1 \sigma_2 \tilde{\sigma}_2\cdots
     \sigma_N \tilde{\sigma}_N\rangle 
\end{gather}
fullfilling $\rho_{\textrm{eq}}(\beta=0) = \textrm{Tr}_Q
|\psi_{\beta=0} \rangle \langle \psi_{\beta=0}| = \hat I / d^N$,
where $d$ is the dimension of the local Hilbert space and $N$ is
the number of physical DoFs. 
The state $|\psi_{\beta=0}\rangle$ can be represented as an MPS
with maximum bond dimension $M = d$ (or as a matrix product
operator with $M = 1$).~\cite{ren2022time}
The thermal state at finite temperature is then obtained via
imaginary-time evolution from $\tau = 0$ to $\tau= \beta/2$. 
\begin{gather}
    |\psi_\beta\rangle = \frac{e^{-\beta \hat H/2} |\psi_{\beta=0}\rangle}
    {\sqrt{\langle \psi_{\beta=0}|e^{-\beta \hat H/2} e^{-\beta \hat H/2}
    |\psi_{\beta=0}\rangle}}
\end{gather}
The desired TCF is obtained from subsequent real-time evolution.
\begin{gather}
    \langle \hat A(t) \hat B\rangle = \frac{\langle
    \psi_{\beta=0}|e^{-\beta \hat H/2} 
    \hat A(t)
    \hat B e^{-\beta \hat H/2} |\psi_{\beta=0}\rangle }{\langle \psi_{\beta=0}|
    e^{-\beta \hat H/2 }  e^{-\beta \hat H/2 } |\psi_{\beta=0}
    \rangle}. \label{eq:purified}
\end{gather}
With static disorder, Eq.~\eqref{eq:purified} becomes 
\begin{gather}
    \langle \hat A(t) \hat B\rangle = \langle \chi| \frac{\langle
    \psi_{\beta=0}|e^{-\beta \hat H(\hat {\mathbf s})/2}
    \hat A(\hat {\mathbf s}, t)
    \hat B(\hat {\mathbf s}) e^{-\beta \hat H(\hat {\mathbf s})/2} |\psi_{\beta=0}\rangle }{\langle \psi_{\beta=0}|
    e^{-\beta \hat H(\hat {\mathbf s})/2 }  e^{-\beta \hat H(\hat {\mathbf s})/2 } |\psi_{\beta=0}
    \rangle} |\chi \rangle \label{eq:FT2}
\end{gather}
This cannot be evaluated by simply evolving $|\psi_{\beta=0}\rangle \otimes |\chi\rangle$, which would yield an incorrect result.
\begin{gather}
   \frac{ \langle \chi|\langle
    \psi_{\beta=0}|e^{-\beta \hat H(\hat {\mathbf s})/2}
    \hat A(\hat {\mathbf s},t)
    \hat B(\hat {\mathbf s}) e^{-\beta \hat H(\hat {\mathbf s})/2} |\psi_{\beta=0}\rangle |\chi \rangle}
    {\langle \chi|\langle \psi_{\beta=0}|
    e^{-\beta \hat H(\hat {\mathbf s})/2 }  e^{-\beta \hat H(\hat {\mathbf s})/2 } |\psi_{\beta=0}
    \rangle |\chi \rangle} \neq \langle \hat A(t) \hat B\rangle  
\end{gather}

\begin{figure}
    \centering
    \includegraphics[width=0.5 \linewidth]{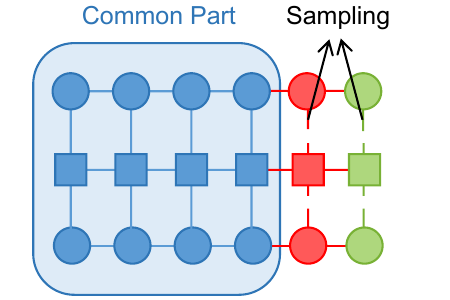}
    \caption{
        Contraction diagram of $\langle \Psi_L(\mathbf{s}_k)|
        \hat{A}(\mathbf{s}_k) | \Psi_R(\mathbf{s}_k) \rangle$
        using matrix product states (MPS) and a matrix product
        operator (MPO). The blue part represents the degrees
        of freedom shared across all samples $\mathbf{s}_k$ and
        can be reused. The red and green part corresponds to
        the auxiliary degrees of freedom associated with static
        disorder, which are sampled to generate $\mathbf{s}_k$.
        The calculation of $\langle \Psi_\beta(\mathbf{s}_k) |
        \Psi_\beta(\mathbf{s}_k) \rangle$ follows a similar
        contraction structure.
       }
    \label{fig:scheme}
\end{figure}

To address this, we calculate three intermediate states, each of
them is represented as an MPS,
\begin{gather}
    |\Psi_\beta \rangle = e^{-\beta \hat H(\hat{\mathbf{s}})/2}
    | \psi_{\beta=0}\rangle |\chi \rangle  \\
    |\Psi_L \rangle = e^{-i \hat H(\hat{\mathbf{s}}) t}
    |\Psi_\beta \rangle  \\
    |\Psi_R \rangle = e^{-i \hat H(\hat{\mathbf{s}}) t} \hat
    B(\hat {\mathbf s}) |\Psi_\beta \rangle 
\end{gather}
Then, according to the discretised distribution function
$\rho(\mathbf s)$,
we sample a set of configurations ${\mathbf{s}_k}$. For each
$\mathbf{s}_k$, we extract:
$|\Psi_\beta(\mathbf s_k)\rangle$, $|\Psi_L(\mathbf s_k)\rangle$,
$|\Psi_R(\mathbf s_k)\rangle$ and $\hat A(\mathbf s_k)$.
Eq.~\eqref{eq:FT2} is then computed as:
\begin{gather}
    \langle \hat A(t) \hat B\rangle = \frac{1}{N_\textrm{samp}} \sum_{\mathbf s_k} \frac{\langle
    \Psi_L(\mathbf s_k)| \hat A (\mathbf s_k)|\Psi_R(\mathbf s_k)
    \rangle}{\langle \Psi_\beta(\mathbf s_k)| \Psi_\beta(\mathbf
    s_k)\rangle} \label{eq:mps_sampling}
\end{gather}
We call this the auxiliary-DoF-based MPS-sampling method.
Each term of Eq.~\eqref{eq:mps_sampling} can be computed via
efficient tensor contractions.
Since for different $\mathbf{s}_k$, the contraction is largely
the same except the sites corresponding to static disorder,
intermediate contractions can be cached and reused at each time
step (as illustrated in Figure.~\ref{fig:scheme}). 
In addition, the denominator (partition function) is
time-independent and thus computed only once.

Although this method involves sampling, it differs from direct
sampling. All sampled terms are extracted from a single
time-evolved MPS that includes auxiliary DoFs, rather than from
many independent simulations. This can significantly reduce
computational cost.
As in the previous subsection, the effectiveness of the method
depends on the similarity of wavefunctions for different
$\mathbf{s}_k$. Encouragingly, in our numerical tests, the MPS
representation remains efficient when the disorder strength is
not excessively large.


\section{Results and discussion} \label{sec:Results}

We benchmark the auxiliary-DoF-based MPS-sampling method by
calculating the dipole-dipole time correlation function,
$\langle \hat{\mu}(t) \hat{\mu} \rangle$, for the emission
spectrum of a one-dimensional Holstein model with static
disorder at finite temperature. The Holstein model is widely
used to describe excited-state dynamics and spectroscopy in
molecular aggregates and materials.~\cite{holstein1959studies,
li2021general,hestand2018expanded,shuai2025excited}
The Hamiltonian including diagonal static disorder is 
\begin{align}
    \hat H & = \sum_i (\varepsilon_0 + s_i)|\phi_i
    \rangle\langle \phi_i| \nonumber \\
    & + \sum_i J(|\phi_i
    \rangle\langle\phi_{i+1}|+ |\phi_{i+1}\rangle\langle\phi_{i}
    |) \nonumber \\ 
     & + \sum_{i} g \omega (b^\dagger_i + b_i)
    |\phi_i \rangle \langle\phi_i| + \sum_i \omega b^\dagger_i b_i
    \label{eq:holstein}
\end{align}
Here, the disorder of site energy $s_i$ is assumed to follow a
Gaussian distribution $s_i \sim \mathcal{N}(0, \sigma^2)$. The
parameters are: vibrational frequency $\omega = 1400\, \text{cm}
^{-1}$, electron-vibration coupling constant $g = 1$, excitonic
coupling $J = -100\, \text{meV}$, monomer transition dipole
moment $\mu = 1$, and temperature $T = 300\, \text{K}$. The
number of harmonic oscillator basis functions per mode is 7, and
time evolution is performed with a 1 fs timestep. The number of
electronic states is denoted as $N_{\text{mol}}$. All
simulations are performed with the Renormalizer
package.~\cite{Renormalizer}

\begin{figure}[htbp]
    \centering
    \includegraphics[width=0.9\linewidth]{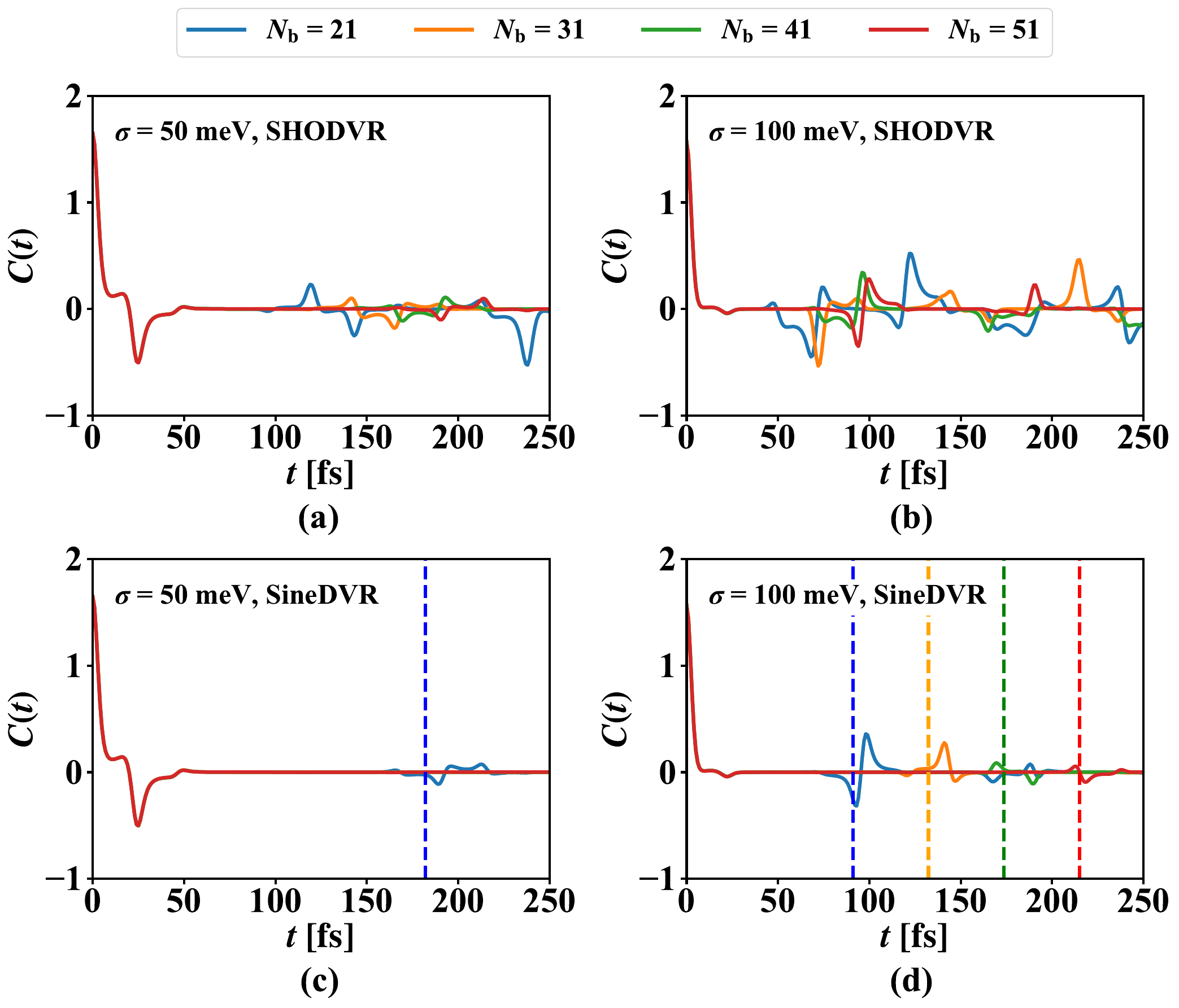}
    \caption{
        Time correlation function $C(t) = \langle \hat{\mu}(t)
        \hat{\mu} \rangle$ for the emission spectrum of a 2-site
        Holstein model at 300 K with varying strengths of static
        disorder (following a Gaussian distribution $\mathcal{N}
        \sim (0, \sigma^2)$) and two types of DVR grids,
        computed using the MPS-sampling method ($M=64$,
        $N_\textrm{samp}=10^6$).
        (a) $\sigma = 50 \, \textrm{meV}$ with SHODVR.
        (b) $\sigma = 100 \, \textrm{meV}$ with SHODVR.
        (c) $\sigma = 50 \, \textrm{meV}$ with SineDVR.
        (d) $\sigma = 100 \, \textrm{meV}$ with SineDVR.
        Different colors represent different numbers of DVR grid
        points $N_\textrm{b}$. The vertical dashed line
        indicates the approximate Poincaré recurrence time, $T =
        \frac{2\pi} {\Delta s}$, for equally spaced SineDVR
        grids.
       }
    \label{fig:basis}
\end{figure}


First, we compare two discrete variable representations of the
auxiliary DoFs: the simple harmonic oscillator DVR (SHODVR) and
sine DVR (SineDVR).~\cite{beck2000multiconfiguration} SHODVR is expected to have
similar accuracy to the bosonic Fock basis used in prior
work.~\cite{gelin2021efficient,sheng2024td} The number of DVR basis functions is denoted as
$N_\text{b}$.
Static disorder is continuous and ideally requires an infinite
basis set, but in practice, we truncate SineDVR to $[-5\sigma,
5\sigma]$, introducing negligible truncation error ($ 5.74 \times
10^{-7}$). For SHODVR, the oscillator frequency is set to
$\omega = 1/(2\sigma^2)$, aligning the ground-state density with
the static disorder distribution. The SineDVR grid points are
uniformly spaced: $s_k^\text{SineDVR} = -5\sigma + k \cdot
\frac{10\sigma}{N_\text{b} + 1},$ while SHODVR grid points are
defined via Hermite polynomial roots: $s_k^\text{SHODVR} =
\sqrt{2}\sigma \cdot \text{root}_k(H_{N_\text{b}}(s))$.

Figure~\ref{fig:basis} compares TCFs for a dimer ($N_\text{mol}
= 2$) at $\sigma = 50$ and $100\, \text{meV}$, using SHODVR and
SineDVR. As expected, both DVRs converge with increasing
$N_\text{b}$. The static-disorder-averaged TCF decays due to
destructive interference across different disorder
configurations $\mathbf{s}_k$. However, with finite basis size,
artificial Poincaré recurrences emerge. Since recurrence time is
inversely related to grid spacing, finer grids delay this
artifact.
SineDVR outperforms SHODVR, showing later and weaker recurrences
with the same grid size, because the SHODVR grid is overly
sparse due to the wide spread of basis functions with $\omega =
1/(2\sigma^2)$.  Therefore, it is preferred to use SineDVR to make
the basis size as small as possible for reducing computational
cost. 
In addition, The uniform spacing $\Delta s = 10\sigma/(N_\text{b}
+1)$ of SineDVR enables estimating the recurrence time as $T
\approx \frac{2\pi} {\Delta s},$ which agrees well with
numerically observed values (vertical dashed lines in
Figure~\ref{fig:basis} (c)(d)). 
This predictive capability offers a practical advantage that the
size of $N_\textrm{b}$ can be appropriately chose according to
the required TCF time range. 
Because of these advantages, SineDVR is recommended for
discretizing the auxiliary DoF and is used in all subsequent
simulations.

\begin{figure}[htbp]
    \centering
    \includegraphics[width=0.99\linewidth]{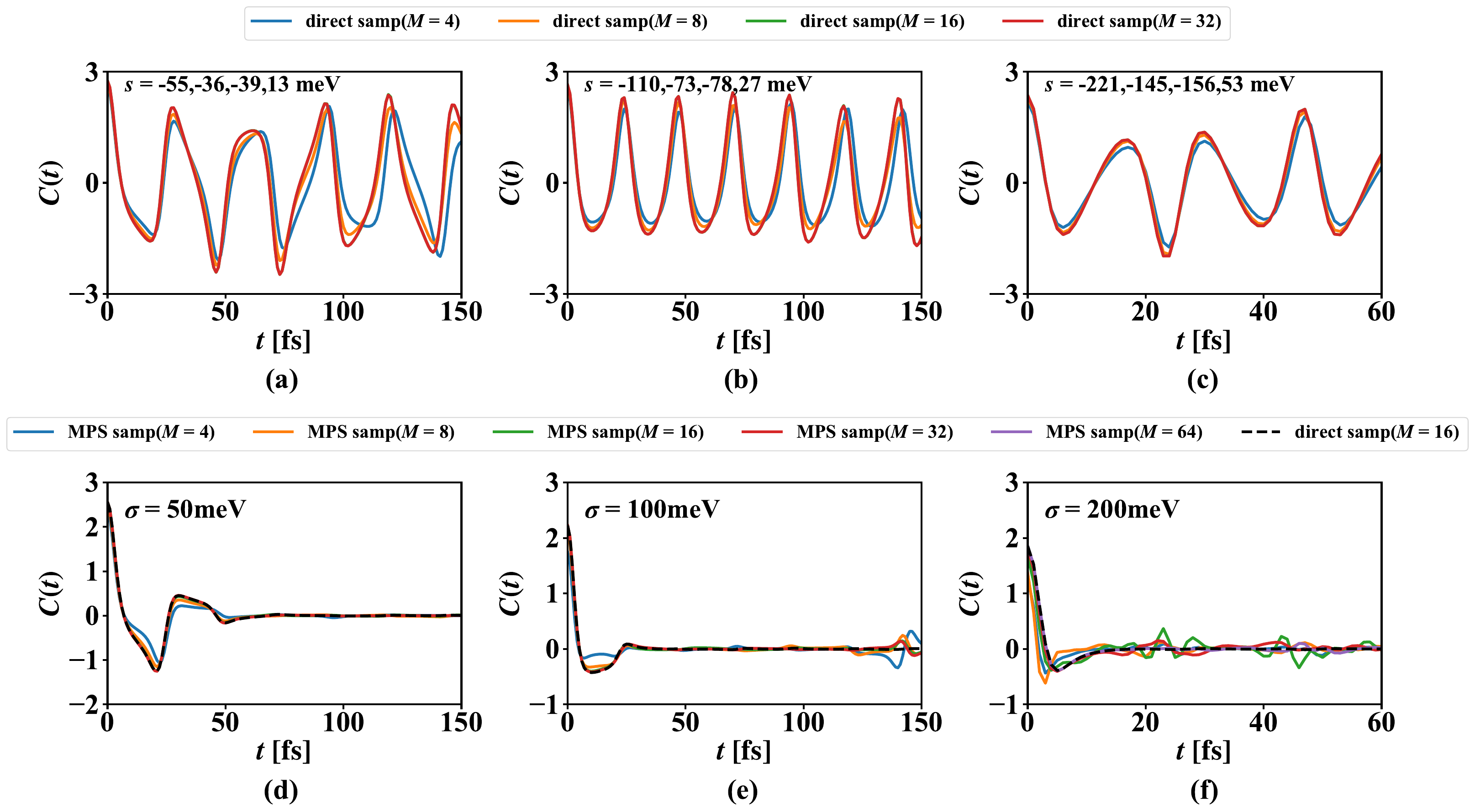}
    \caption{
        Time correlation function $C(t) = \langle \hat{\mu}(t)
        \hat{\mu} \rangle$ for the emission spectrum of a 4-site
        Holstein model at 300 K with different bond dimensions
        $M$.  (a-c) Results from direct sampling for a single
        realization with $\sigma = 50 \, \textrm{meV}$, $100 \,
        \textrm{meV}$, and $200 \, \textrm{meV}$, respectively.
        The static disorder values $s_i$ for each site are
        labeled in the figures. Different colors represent
        different bond dimensions.  (d-f) Results from the
        MPS-sampling method. The number of DVR grid points is
        $N_\textrm{b} = 21$ for $\sigma = 50 \, \textrm{meV}$ and
        $N_\textrm{b} = 31$ for $\sigma = 100 \, \textrm{meV}$ and
        $200 \, \textrm{meV}$. The black dashed lines indicate the
        reference results from direct sampling. Both methods use
        $N_\textrm{samp} = 2 \times 10^4$.
       }
    \label{fig:M}
\end{figure}

The next key question is whether the auxiliary DoFs significantly
increase the MPS bond dimension $M$, or in other words, whether
the total wavefunction $\Psi(t)$ in Eq.~\eqref{eq:linearcombine}
is efficiently compressible. The computational cost of TD-DMRG
with the projector-splitting time evolution algorithm  scales as
$\mathcal{O}(M^n)$ with $n \approx 2 \sim 3$.~\cite{li2020numerical,haegeman2016unifying} If
$M$ becomes too large, it would offset the efficiency gains of
avoiding multiple independent realizations.
Figure~\ref{fig:M}(a-c) show TCFs for three random disorder
realizations of a 4-site Holstein model, at $\sigma = 50$, $100$,
and $200\, \text{meV}$, respectively. In all cases, a bond
dimension of $M = 16$ suffices for convergence. Figure~\ref{fig:M}(d-f) 
compare the MPS-sampling method with
the direct sampling method (same number of samples, $N_\text{samp} =
2 \times 10^4$). The agreement validates the new method.
For $\sigma = 50$ and $100\, \text{meV}$, the required $M = 16$
is similar to that of single realizations, and thus the overall
cost reduction is roughly proportional to $N_\text{samp}$, since
overhead in sampling MPS is cheap due to efficient tensor
contractions (see data below). 
For $\sigma = 200 \, \text{meV}$, a larger bond dimension ($M =
64$) is required due to greater variation among wavefunctions
corresponding to different disorder strengths. As the static
disorder strength approaches the electronic bandwidth
(approximately $320 \, \text{meV}$ for the 4-site model with $J =
-100 \, \text{meV}$), the system undergoes a transition from
delocalized to localized states, a phenomenon known as Anderson
localization.~\cite{anderson1958absence} In this case,
compressing both delocalized and localized wavefunctions into a
single MPS, as in Eq.~\eqref{eq:linearcombine},
becomes inefficient. Therefore, the auxiliary-DoF-based method
is most effective when the disorder strength remains smaller
than the electronic bandwidth.

\begin{figure}[htbp]
    \centering
    \includegraphics[width=0.9\linewidth]{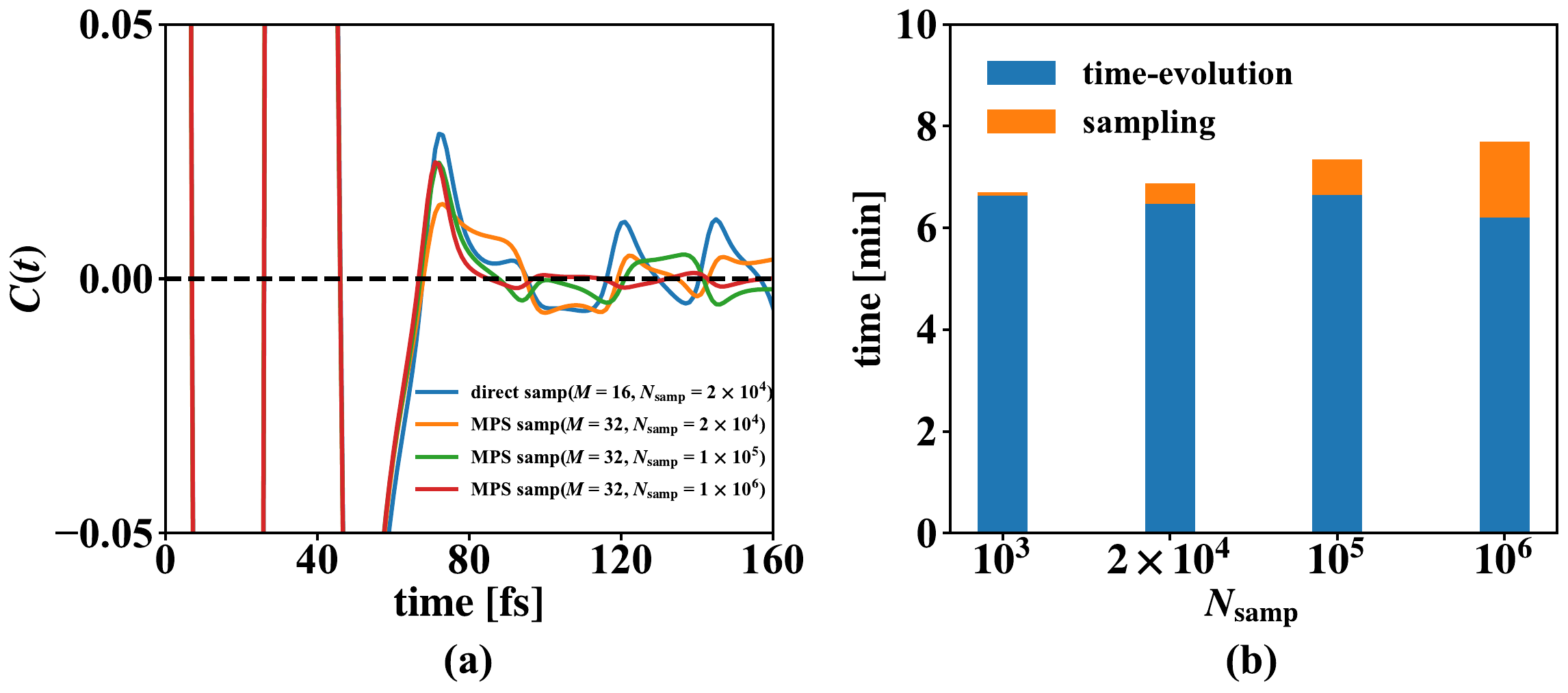}
    \caption{
        Time correlation function $C(t) = \langle \hat{\mu}(t)
        \hat{\mu} \rangle$ (a) and wall time (b) for computing
        the emission spectrum of a 4-site Holstein model at 300
        K using the MPS-sampling method with varying sample
        sizes $N_\textrm{samp}$. The static disorder strength is
        $\sigma = 50 \, \textrm{meV}$, and the number of DVR grid
        points per auxiliary degree of freedom is $N_\textrm{b}
        = 21$. Wall time measurements were performed over 10
        time steps using 4 EPYC 7B13 CPU cores and 1 NVIDIA A100
        GPU.
       }
    \label{fig:sample_error}
\end{figure}

Since MPS-sampling involves efficient tensor contractions and
benefits from reuse of intermediates, it enables much larger
sample sizes with minimal overhead than the direct sampling
approach to reduce the statistical error. Figure~\ref{fig:sample_error} (a) shows reduced noise in the TCF of 4-site model
as $N_\text{samp} $ increases from $2\times 10^4$ to $10^6$.
Figure~\ref{fig:sample_error}(b) plots the corresponding wall
time for evolving 10 steps.
Although the sampling time (i.e., computing the numerator and
denominator of Eq.~\eqref{eq:mps_sampling}) increases with
sample size, it remains a small fraction of the total cost
compared to time evolution. As a result, the overall runtime
grows only modestly with increasing sample size.

\begin{figure}[htbp]
    \centering
    \includegraphics[width=0.5\linewidth]{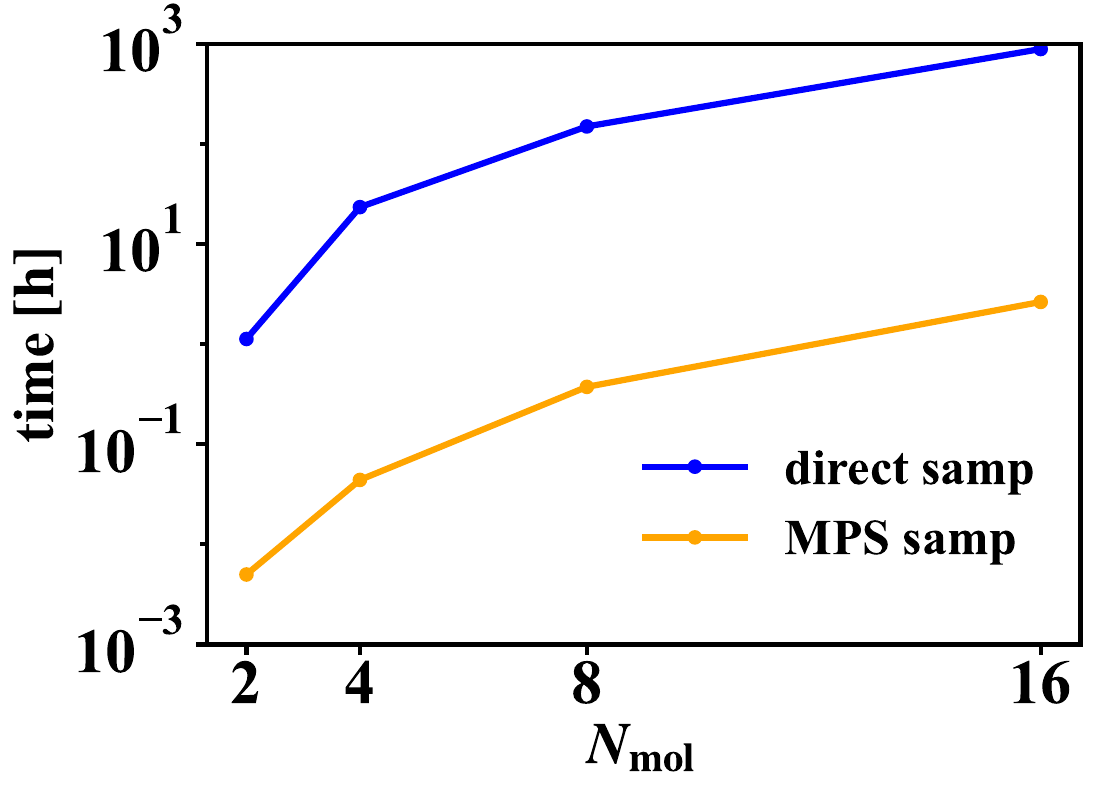}
    \caption{
        Wall time for computing the dipole-dipole time
        correlation function of the Holstein model at 300 K
        using the MPS-sampling method (yellow) and direct
        sampling method (blue) for various system sizes
        $N_\textrm{mol}$. Parameters: $\sigma = 50 \, \textrm{meV}
        $, $N_\textrm{b} = 21$, and $N_\textrm{samp} = 3000$.
        Bond dimensions used are: $M = 4$ (direct sampling) and
        $M = 8$ (MPS-sampling) for $N_\textrm{mol} = 2$; $M =
        16$ for $N_\textrm{mol} = 4$; $M = 32$ for
        $N_\textrm{mol} = 8$; and $M = 48$ for $N_\textrm{mol} =
        16$.
       }
    \label{fig:time}
\end{figure}

Finally, Figure~\ref{fig:time} compares the wall times of direct
sampling and MPS-sampling methods across various system sizes,
using $N_\text{samp} = 3000$, a typical value for capturing
static disorder effects. The bond dimensions used in each
converged simulation are listed in the caption. The MPS-sampling
method achieves a speedup of about two orders of magnitude,
clearly demonstrating its computational advantage, especially as
system size increases.

\section{Conclusion} \label{sec:Conclusion}

In conclusion, we have developed an MPS-sampling method that
extends the auxiliary degree-of-freedom approach to efficiently
treat static-disorder-dependent thermal equilibrium initial
states in quantum dynamics simulations. This extension enables
the accurate computation of equilibrium time correlation
functions in disordered systems through a one-shot simulation,
eliminating the need for extensive ensemble averaging.

Our method captures the effects of static disorder with only a
moderate increase in MPS bond dimension compared to conventional
direct sampling. It is particularly effective when the disorder
strength is smaller than the electronic bandwidth, as
demonstrated in benchmarks using the Holstein model. Even for
stronger disorder approaching the bandwidth, the required bond
dimension increases but remains tractable. 
We also examined different DVR grid choices for discretizing the
auxiliary DoF and found that the sine DVR offers superior
convergence and lower recurrence artifacts compared to the
harmonic oscillator DVR, making it the preferred basis for
practical simulations.

Overall, the MPS-sampling approach offers substantial
computational savings, achieving up to a two-order-of-magnitude
speedup over traditional methods that require thousands of
independent realizations. Furthermore, it enables the efficient
generation of very large sample sizes (exceeding $10^6$) with
minimal additional cost, significantly reducing statistical
noise. These advantages make the method a powerful and scalable
tool for studying static disorder effects in complex quantum
systems.

\section{Acknowledgments}

    This work is supported by the National Natural Science
    Foundation of China (Grant No. 22273005 and No. 22422301),
    the Innovation Program for Quantum Science and Technology
    (Grant No. 2023ZD0300200), NSAF (Grant No. U2330201), and
    the Fundamental Research Funds for the Central Universities.

\bibliography{ref}

\end{document}